\documentclass[preprint,epsfig,floats,aps]{revtex4}
\usepackage{epsfig}
\begin{document}
%\baselinestretch{5}
%\tightenlines
\title{Energy dependence of string fragmentation function\\
       and $\phi$ meson production}
%\vspace{0.1in}
\author{Ben-Hao Sa$^{1,2,4,5}$\footnotemark
 \footnotetext{E-mail: sabh@iris.ciae.ac.cn},Xu Cai$^{4,2}$,
  Chinorat Kobdaj$^3$, Zhong-Qi Wang$^1$, Yu-Peng Yan$^3$, 
  and Dai-Mei Zhou$^2$
 }
\affiliation{
$^1$  China Institute of Atomic Energy, P. O. Box 275 (18),
      Beijing, 102413 China \\
$^2$  Institute of Particle Physics, Huazhong Normal University,
      Wuhan, 430079 China \\
$^3$  School of Physics, Suranaree University of Technology,
       Nakhon Ratchasima 30000, Thailand\\
$^4$  CCAST (World Lab.), P. O. Box 8730 Beijing, 100080 China\\
$^5$  Institute of Theoretical Physics, Academia Sinica, Beijing,
      100080 China
}
%\maketitle
\begin{abstract}
The $\phi$ meson productions in $Au+Au$ and/or $Pb+Pb$ collisions at AGS, SPS,
RHIC, and LHC energies have been studied systematically with a hadron and 
string cascade model LUCIAE. After considering the energy dependence of the 
model parameter $\alpha$ in string fragmentation function and adjusting it 
to the experimental data of charged multiplicity to a certain extent,
the model predictions for $\phi$ meson yield, rapidity, and/or
transverse mass distributions are compatible with the experimental data at 
AGS, SPS and RHIC energies. A calculation for $Pb+Pb$ collisions at LHC energy 
is given as well. The obtained fractional variable in string fragmentation 
function shows a saturation in energy dependence. It is discussed that the 
saturation of fractional variable in string fragmentation function might be a 
qualitative representation of the energy dependence of nuclear transparency.
\end{abstract}
\vspace{0.1in}
\maketitle

Strangeness enhancement was suggested in the early eighties \cite{raf1} as one 
of the most promising signatures for the creation of a Quark-Gluon Plasma
(QGP) phase in relativistic nuclear collisions. Following the experimental 
observations on strangeness enhancement in proton-nucleus and nucleus-nucleus 
collisions at the SPS energies \cite{na35,na36,wa85,wa94,na44,na38}  
the WA97 has measured a clear enhancement of multi-strange baryons ($\Lambda, 
\Xi, \Omega$) with their strange quark content in 158A GeV/c $Pb+Pb$ 
collisions relative to $p+Pb$ collisions \cite{wa97}. Recently the STAR data 
on the strangeness production in $Au+Au$ collisions at $\sqrt{s_{nn}}$=130 GeV 
were reported \cite{star1}. 

\small{
\begin{table}[hb]
\caption{Global hadron multiplicity in
                   relativistic nucleus-nucleus collisions}
%\begin{center}
\begin{tabular}{|c|c|c|c|c|c|}
\hline
\hline
\multicolumn{2}{|c|}{Reaction} &$Au+Au$ &$Pb+Pb$ &$Au+Au$ &$Pb+Pb$ \\
\hline
\multicolumn{2}{|c|}{$\sqrt{s_{nn}}$ (GeV)} &4.66 &17.3 &130 &5500 \\
\hline
Centrality &Exp. &$\leq 4\%$ &$\leq 4\%$ &$\leq 5\%$ (BRAHMS) &$\leq 10\%$ \\
\cline{2-6}
&LUCIAE &$b\leq$2.82 fm &$b\leq$3.5 fm &$b\leq $3.2 fm &$b\leq$4.54 fm \\
\hline
$N_{ch}$ &Exp. & & &3860$\pm$300$^{1)}$ & \\
\cline{2-6}
&LUCIAE & & &4114 &15646$^{2)}$ \\
\hline
$(<\pi^+>+<\pi^->)/2$ &Exp. & 105$^{3)}$ &611$^{4)}$ & & \\
\cline{2-6}
&LUCIAE &102 &611 & &5851$^{2)}$ \\
\hline
\multicolumn{2}{|c|}{$\alpha$} &0.4 &1.3 &12 &18 \\
\hline
\multicolumn{2}{|c|}{$<z>$} &0.286 &0.565 &0.923 &0.947 \\
\hline
\hline
\multicolumn{3}{l}{1. -4.7$<\eta<$4.7, from \cite{brah}} &\multicolumn{3}{l}
 {2. in full phase space}\\
\multicolumn{3}{l}{3. $0.6<y<2.6$, taken from \cite{ahle}} &\multicolumn{3}
{l}{4. cf. \cite{na49}}\\
\end{tabular}
%\end{center}
\end{table}
} 
   
\begin{figure}[ht]
\centerline{\hspace{-0.5in}
\epsfig{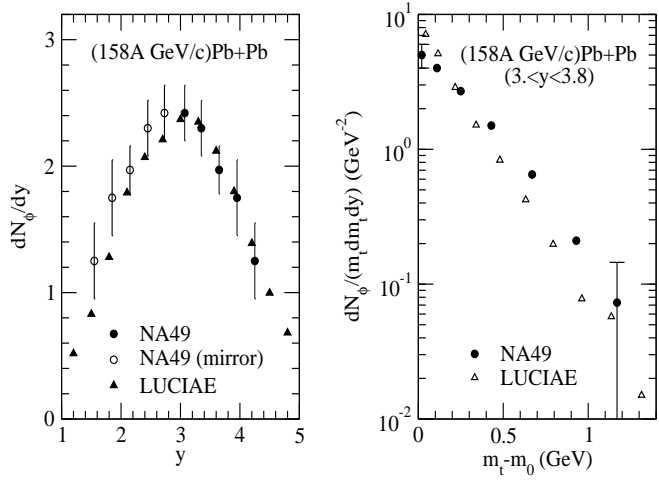}}
\vspace{0.2in}
\caption{Rapidity (left panel) and transverse mass (right panel) 
 distributions of $\phi$ mesons in $Pb+Pb$ collisions at 158A GeV/c}
\label{pbpb_158}
\end{figure}

As the mesonic counterpart, the enhancement of $\phi$ meson production 
in relativistic nuclear collisions was also suggested as an evidence of the 
QGP formation in Ref. \cite{shor}, since in the environment of a QGP the 
copious strange and antistrange quarks originating from gluon annihilation 
would be very likely to coalesce forming $\phi$ mesons during the 
hadronization period. Due to the small cross sections of $\phi$ mesons 
interacting with non-strange hadrons \cite{raf1,shor}, penetrating $\phi$ 
mesons are also messengers of the early stage of the colliding system. Thus, 
the $\phi$ meson is not only a promising signature for the QGP formation but 
also a good probe to study the reaction dynamics. After the experimental 
observations on $\phi$ meson productions in $Au+Au$ collisions at AGS 
\cite{e917} and sulfur-nucleus collisions and $Pb+Pb$ collisions at SPS 
\cite{na382,helios,na49}, the STAR collaboration reported recently the data on 
$\phi$ meson productions in $Au+Au$ collisions at $\sqrt{s_{nn}}$=130 GeV 
\cite{star2}. 

\begin{figure}[ht]
\centerline{\hspace{-0.5in}
\epsfig{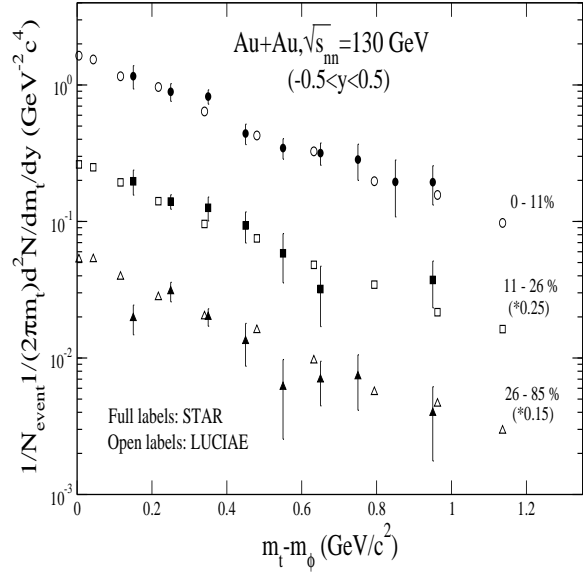}}
\vspace{0.2in}
\caption{Transverse mass distributions (-0.5$<y<$0.5) of 
 $\phi$ mesons in $Au+Au$ collisions at $\sqrt{s_{nn}}$=130 GeV}
\label{mtau_130}
\end{figure}

The model studies on the $\phi$ meson enhancement in relativistic 
nucleus-nucleus collisions are rare \cite{ko1,bere,sa3,pal} and there exists 
up to now no theoretical investigation on the energy dependence of $\phi$ 
production from AGS, to SPS, to RHIC, and up to LHC energies systematically. 
In this letter, a hadron and string cascade model, LUCIAE \cite{sa} was 
employed to investigate first time the energy dependence of $\phi$ meson 
production systematically. We have successfully used LUCIAE to study the 
enhanced production of multi-strange baryons ($\Lambda, \Xi, \Omega$) and 
determined the model parameters related to the production of strange particles 
\cite{tai1,sa2}.  

\small{
\begin{table}[hb]
\caption{$\phi$ mesons yield in relativistic 
                   nucleus-nucleus collisions}
\begin{tabular}{|c|c|c|c|c|c|}
\hline
\hline
Reaction &$\sqrt{(s_{nn})}$ &\multicolumn{2}{|c|}{Centrality} &
 \multicolumn{2}{|c|}{$<N_{\phi}>$} \\
\cline{3-6}
 &(GeV) &Exp. &LUCIAE &Exp. data &LUCIAE \\
\hline
$Au+Au$ &4.66 &301$^{1)}$ &307$^{1)}$ &0.252$\pm$0.107$^{2)}$& 0.14 \\
\hline
$Pb+Pb$ &17.3 &$\leq 4\%$ &$b\leq$3.5 fm$^{3)}$ &7.6$\pm 1.1^{3)}$ &6.48 \\
\hline
$Au+Au$ &130  &$\leq 11\%$ (STAR) &$b\leq$4.31 fm &5.73$\pm$1.06$^{4)}$ &5.10\\
\hline
$Pb+Pb$ &5500 &$\leq 10\%$ &$b\leq$4.54 fm & &132$^{5)}$ \\
\hline
\hline
\multicolumn{3}{l}{1. $<N_{part}>$} &
\multicolumn{3}{l}{2. $\frac{dN_{\phi}}{dy}$ within 0.9$<y<$1.4 \cite{e917}}\\
\multicolumn{3}{l}{3. cf. \cite{na49}} &
\multicolumn{3}{l}{4. $\frac{dN_{\phi}}{dy}$ within -0.5$<y<$0.5 \cite{star2}}\\
\multicolumn{6}{l}{5. full phase space}
\end{tabular}
\end{table}
}

The LUCIAE model is based on FRITIOF \cite{pi}, which is an incoherent hadron 
multiple scattering and string fragmentation model. In FRITIOF, the nucleus-
nucleus collision is depicted simply as a superposition of nucleon-nucleon 
collisions. What characterizes LUCIAE beyond FRITIOF are the following 
features: First of all, the rescattering among the participant and spectator 
nucleons and the produced particles from string fragmentation are generally 
taken into account \cite{sa1}. However, as proposed in \cite{raf1,shor} the 
effects of the final state interactions on the $\phi$ meson production and 
propagation are neglected. Secondly, the collective effect in the gluon 
emission of strings is considered by firecracker model \cite{tai}. 
Thirdly, a phenomenological mechanism for the reduction of the $s$ quark 
suppression in the string fragmentation is introduced \cite{tai1} 
resulting the effective string tension and therefore the pertained JETSET 
parameters vary automatically with the energy, centrality, and size of 
reaction system. Those JETSET parameters: parj(2), (3), (1), and (21) in 
program are, respectively, the suppression of s quark pair production in 
string fragmentation compared with u or d pair production, the extra 
suppression of strange diquark production compared with the normal suppression 
of strange quarks, the suppression of diquark-antidiquark pair production 
compared with quark-antiquark pair production, and the width in the Gaussian 
distribution of the transverse momentum of primary hadron \cite{sjo1}.      

\begin{figure}[ht]
\centerline{\hspace{-0.5in}
\epsfig{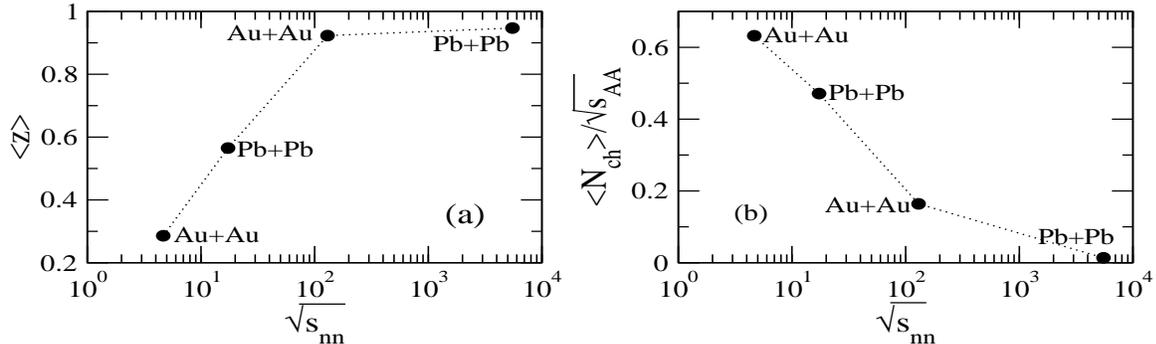}}
\vspace{0.2in}
\caption{(a) fractional variable in string fragmentation function $<z>$   
 and (b) $<N_{CH}>$/$\sqrt{s_{AA}}$ as a function of $\sqrt{s_{nn}}$}
\label{avech}
\end{figure}

From the AGS energy ($\sqrt{s_{nn}}\sim$ 5 GeV) to SPS ($\sqrt{s_{nn}}\sim$ 
20 GeV) to RHIC ($\sqrt{s_{nn}}\sim$ 200 GeV) and then to the LHC energy 
($\sqrt{s_{nn}}\sim$ 5500 GeV) there are four energy magnitudes spanned. Among 
the nucleus-nucleus collisions in above energy region their dynamic behaviors, 
of course, must be different observably from each other. To describe the 
nucleus-nucleus collisions in such a wide energy region one either selects 
suitable model for different energy or considers the energy dependence of 
model ingredients properly if one single model is used. This idea has indeed 
been adopted in $e^+e^-$ studies quite early \cite{ochs}. Recently the thermal 
model with two main parameters (temperature $T$ and  baryon chemical 
potential $\mu_B$) was used to fit the experimental data of global hadronic 
yields for nucleus-nucleus collisions at SIS ($\sqrt{s_{nn}}\sim$ 1 GeV), AGS, 
SPS, and RHIC energies, the obtained values of $T$ and $\mu_B$ are quite 
different from each other \cite{cley}. Therefore, for dynamical models it is 
also needed to consider the energy dependence of the involved dynamical 
ingredients properly in order to describe well the nucleus-nucleus collisions 
in region spanned four energy magnitudes. Of course, some dynamical 
ingredients might be sensitive to the energy and others might not be. Many 
dynamical models (including LUCIAE) working well for nucleus-nucleus 
collisions at SPS energy, however, overestimate the charged multiplicity at 
RHIC energy. One reason might just be that there are ingredients either 
energy independent or energy dependent without taken into account properly.  

In LUCIAE model (in FRITIOF and in JETSET eventually) a very important 
dynamical ingredient, the string fragmentation function, is energy 
independent. Thus the LUCIAE model is extended in this letter to take the 
energy dependence of string fragmentation function into account. As mentioned 
in \cite{sjo1} that the string fragmentation function $f(z)$, which expresses 
the probability that a given fractional variable $z$ is sampled, could be  
arbitrary in principle. In \cite{sjo1} several such choices are given, besides 
the one of Lund fragmentation function. For describing easily the relativistic 
nucleus-nucleus collisions over four energy magnitudes one of the simplest 
option for the string fragmentation function in \cite{sjo1} is employed here 
for the moment. That string fragmentation function reads 
 \begin{equation}
 f(z)=\alpha {(\sqrt{s_{nn}})} \times z^{\alpha{(\sqrt{s_{nn}})}-1},
 \end{equation} 
where the fractional variable $z$ refers to the fraction of light-cone 
momentum taken away by produced hadron from the fragmenting string, the 
parameter $\alpha$ in this string fragmentation function is assumed to be 
energy dependent now.

We first ran LUCIAE for $Pb+Pb$ and $Au+Au$ collisions, respectively, at SPS 
and RHIC energies. In LUCIAE calculations the centrality cut and the windows 
of rapidity (pseudorapidity) and $p_t$ were the same as those in corresponding 
experiments (the same later). Adjusting the parameter $\alpha{(\sqrt{s_{nn}})}
$ in order that the global hadron multiplicity from LUCIAE could be 
compatible with the corresponding experimental data, the obtained LUCIAE 
results are given in Tab. 1 together with the experimental data. The data of 
$Pb+Pb$ collisions at SPS energy were taken from \cite{na49} and $Au+Au$ 
collisions at $\sqrt{s_{nn}}$=130 GeV from \cite{brah}. Since in LUCIAE the 
reduction mechanism of $s$ quark suppression was introduced under the 
requirement that the pertained JETSET parameters mentioned above, must be 
reduced to the corresponding default values for the $p+p$ collisions at SPS 
energy, the LUCIAE model with this mechanism works for SPS energy and above 
only. Therefore we ran LUCIAE without reduction mechanism of $s$ quark 
suppression for $Au+Au$ collisions at AGS energy but a empirical value of $s$ 
quark suppression factor (parj(2)=0.45) extracted from experiments \cite{cley} 
was used in stead of default ones in JETSET. The experimental data for $Au+Au$ 
collisions at AGS energy were taken from \cite{ahle}. The LUCIAE calculation  
for $Pb+Pb$ collisions at LHC energy was also given in Tab. 1, in that   
calculation the $\alpha{(\sqrt{s_{nn}})}$=18, larger properly than that in 
$Au+Au$ collisions at RHIC, was assumed. In Tab. 1 the fixed $\alpha{(\sqrt{
s_{nn}})}$ parameter and the obtained fractional variable in string 
fragmentation function $<z>$ (averaged over events), were given as well. 

Then we employed the parameter $\alpha{(\sqrt{s_{nn}})}$ fixed above to 
calculate the $\phi$ meson yield, rapidity, and/or transverse 
mass distributions in relativistic nucleus-nucleus collisions. The LUCIAE 
results of $\phi$ meson yield were compared with the corresponding data in 
Tab. 2, where the experimental data were taken from \cite{e917}, \cite{na49}, 
and \cite{star2} for $Au+Au$ collisions at AGS energy, $Pb+Pb$ at SPS, and 
$Au+Au$ at $\sqrt{s_{nn}}$=130 GeV, respectively. One sees in Tab. 2 that 
globally speaking the LUCIAE results are well compatible with experimental 
data.

In Fig. 1 we compare the LUCIAE results of $\phi$ meson rapidity (left panel) 
and transverse mass (right panel, 3.0$<$y$<$3.8) distributions with the 
experimental data in $Pb+Pb$ collisions at SPS energy (data taken from 
\cite{na49}). It is shown in Fig. 1 that the LUCIAE model works 
somewhat better for the rapidity distribution than the transverse mass 
distribution in $Pb+Pb$ collisions at SPS energy. The comparison between 
LUCIAE results and experimental data of $\phi$ meson transverse mass 
distributions (-0.5$<$y$<$0.5) in $Au+Au$ collisions at $\sqrt{s_{nn}}$=130 
GeV was given in Fig. 2 where data were taken from \cite{star2}. One sees in  
Fig. 1 and 2 that for transverse mass distribution LUCIAE model works better 
at RHIC energy than SPS since more hard $\phi$ mesons are produced at RHIC 
than SPS due to the Schwinger mechanism \cite{schw} in particle production of 
string fragmentation \cite{sa3}.

Fig. 3 (a) gives the fractional variable in string fragmentation function, 
$<z>$, as a function of $\sqrt{s_{nn}}$ from LUCIAE calculations above. A 
saturation structure at around RHIC energy is observed in Fig. 3 (a). Since  
the fractional variable $z$ is the fraction of light-cone momentum taken away 
from the fragmenting string by the produced hadron, it must relate strongly to 
the nuclear transparency in nucleus-nucleus collisions. The formation time in 
which string is excited after hard scattering and the hadronic rescattering 
were considered as dominative factors in baryon stopping (nuclear transparency
) \cite{urqmd}. However, one knows that the higher reaction energy the longer 
formation time, and then the less rescattering and higher nuclear 
transparency. As for the rescattering itself, the higher reaction energy leads 
to the more produced particles, and then the more rescattering and lower 
nuclear transparency. Thus the effects of formation time and the hadronic 
rescattering on the nuclear transparency cancel each other in certain extent. 
One might expect that the fractional variable in string fragmentation function 
is dominant in nuclear transparency and its saturated energy dependence is a 
qualitative representation of the energy dependence of nuclear transparency. 

The corresponding LUCIAE results of charged multiplicity per unit reaction 
energy $\displaystyle{<N_{ch}>/\sqrt{s_{AA}}}$ in full phase space as a 
function of $\sqrt{s_{nn}}$ are given in Fig. 3 (b). A trend of saturation 
seems to be there in Fig. 3 (b) as well. One sees in Fig. 3 (a) that the 
hadron is created with higher fractional variable as $\sqrt{s_{nn}}$ is 
increased and the fractional variable approaches almost to its maximum value 
one when $\sqrt{s_{nn}}$ gets the LHC energy. On the other hand, Fig. 3 (b) 
denotes that the efficiency rate of energy to hadron (matter) is decreasing 
with increase of $\sqrt{s_{nn}}$ and tends also to some kind of saturation. 
Since the fractional variable equal to one means that only one hadron would 
be created from fragmentation of an exited string and the number of excited 
strings (nucleon-nucleon collisions) is limited, the fractional variable 
approaching to unit must lead to the saturated trend in $\displaystyle{<N_{ch}
>/\sqrt{s_{AA}}}$. Those saturation trends might be an indication of the 
saturation in particle production. 

In summary, we employed a hadron and string cascade model, LUCIAE, to study 
the $\phi$ meson production in $Au+Au$ and/or $Pb+Pb$ collisions at AGS, SPS,
RHIC, and LHC energies. By considering the energy dependence of the model   
parameter $\alpha$ in the string fragmentation function and adjusting it to 
the experimental data of charged multiplicity to a certain extent, the model 
results of $\phi$ meson yield, rapidity, and/or transverse mass distributions 
were compatible with the corresponding experimental data for AGS, SPS and RHIC 
energies. A calculation for the $\phi$ meson production in $Pb+Pb$ collisions 
at LHC energy is also given. The obtained fractional variable in string 
fragmentation function shows a saturation in energy dependence at around RHIC 
energy. It is discussed that this energy saturation phenomenon could be a 
qualitative representation of the energy dependence of nuclear transparency.  
  
Finally, the financial supports from NSFC in China (19975075, 10135030, and 
10075035) and NRCT in Thailand (1.CH7/2454) are acknowledged.

\end{document}